\documentclass[aps,prl,twocolumn,nofootinbib,groupedaddress,amsfonts,floatfix]{revtex4}

\usepackage{graphicx,amsmath,amssymb,amstext}
\usepackage{amssymb,amsbsy,amsfonts,amsthm}

\begin{document}

\title{Gravitational Radiation from First-Order Phase Transitions}
 
\author{Hillary L. Child${}^1$}
%\email[]{richard.easther@yale.edu}
\author{John T. Giblin, Jr${}^{1,2}$}
%\email[]{jgiblin@perimeterinstitute.ca}

\affiliation{${}^1$Department of Physics, Kenyon College, Gambier, OH 43022}
\affiliation{${}^2$Department of Physics, Case Western Reserve University, Cleveland, OH 44106}

\begin{abstract}
It is believed that first-order phase transitions at or around the GUT scale will produce high-frequency gravitational radiation.  This radiation is a consequence of the collisions and coalescence of multiple bubbles during the transition.  We employ high-resolution lattice simulations to numerically evolve a system of bubbles using only scalar fields, track the anisotropic stress during the process and evolve the metric perturbations associated with gravitational radiation.  Although the radiation produced during the bubble collisions has previously been estimated, we find that the coalescence phase enhances this radiation even in the absence of a coupled fluid or turbulence.  We comment on how these simulations scale and propose that the same enhancement should be found at the Electroweak scale; this modification should make direct detection of a first-order electroweak phase transition easier.
\end{abstract}

\maketitle

Gravitational waves are a direct probe of the physics of the early Universe and the only direct probe of physics before recombination.  The direct detection of the stochastic gravitational wave background will carry with it new insight to high energy physics and the nature of gravity.  Cosmological processes that produce stochastic gravitational wave signals---inflation, cosmic string networks,  phase transitions---are of substantial interest since they represent an untapped source of information.  

Direct detection experiments such as the Laser Interferometer Gravitational-wave Observatory (LIGO) \cite{LIGO} are now underway and the hope of future missions, e.g. \cite{LISA}, will further the direct detection search; the time for {\sl a priori} estimation of signatures will soon come to an end.  Our current intention is to provide precision estimates of the gravitational wave spectrum to aid in the design of the next generation of observatories.

The spectrum of gravitational waves from first-order phase transitions has been of interest for a number of decades.  Although estimates of the gravitational radiation from cosmological processes date to the the early 1980's \cite{Witten:1984rs}, the first rigorous attempts at estimating the power radiated from the collisions of bubbles came somewhat later, first from vacuum bubbles \cite{Kosowsky:1991ua,Kosowsky:1992vn} and then in the context of phase transitions \cite{Kosowsky:1992rz,Kamionkowski:1993fg}.  It was in this later work that the authors discovered that turbulence would additionally contribute to the stochastic gravitational wave spectrum. In recent times, interest in the gravitational radiation produced during phase transitions, with particular emphasis on the electroweak phase transition \cite{Caprini:2006jb,Gogoberidze:2007an,Caprini:2007xq,Kahniashvili:2008pf,Caprini:2009fx}, has led to good estimates and has rejuvenated the discussion.  These estimates accurately address the power from {\sl detonations}: highly energetic domain walls colliding at large velocities and a subsequent period of turbulence (the authors of \cite{Huber:2008hg} have additionally addressed the parametric dependence of that terminal velocity to the results of the preceding work).  Analytic work has begun to address the added complexity of {\sl deflagrations}.  In this case thick walls (preceded by shock fronts) smear out the large field gradients, and exact descriptions of these collisions will need to be explored using fluid dynamics.  We leave the topic of deflagrations to a future manuscript.

Although much of this work has also addressed the added contribution of gravitational wave power from turbulence produced by the coupled fluid during the phase transition, we aim to present an additional effect.  Even in the absence of proper fluid turbulence, a coalescent phase of the fields participating in the phase transition amplifies the gravitational wave signal and shift the peak frequency.  Although this is discussed as possible for the electroweak phase transition, see e.g. \cite{Bodeker:2009qy}, we see this as a toy model where we concentrate on the generation of gravity waves from a scalar-only phase transition.

Here, we discuss the possibility that the Universe underwent a first-order phase transition at or near the GUT scale, $\sim 10^{13}-10^{15} \,\rm{GeV}$, at time $t_*$.  Later this time will correspond to a program time, $\tau=0$.  We realize the phase transition with a scalar field, $\phi$, and a Coleman-Weinberg type associated potential,
\begin{equation}
V(\phi) = \frac{\lambda}{8}\left(\phi^2-\phi_0^2\right)^2 + \epsilon \lambda \phi_0^3(\phi+ \phi_0),
\end{equation}
where $\epsilon$ parameterizes the height difference between the two minima, $\lambda$ is the dimensionless self-coupling of the fields and $\phi_0$ sets the scale of the transition.  We parameterize the energy density, $\rho_*$, of the whole Universe at the time of the transition by the energy scale, $\mu$, such that
\begin{equation}
\rho_* = \mu^4.
\end{equation}
Just before the transition begins, the field is static, so the homogeneous energy density of the field is given by the energy of the metastable minimum,
\begin{equation}
\rho_\phi(t_*) = 2 \epsilon \lambda \phi_0^4.
\end{equation}
We presume that the field undergoing the phase transition is some fraction, $\xi$, of the total content of the Universe at that time, so the energy in the field is 
\begin{equation}
\rho_\phi(t_*)=2 \epsilon \lambda \phi_0^4 = \xi \mu^4.
\end{equation}
Most of the literature that studies phase transitions parameterizes energy using the ratio of the energy of the false minimum compared to the overall thermal energy \cite{Kamionkowski:1993fg}, for this case,
\begin{equation}
\alpha = \frac{2\epsilon \lambda \phi_0^4}{(1-\xi)\mu^4} = \frac{\xi}{1-\xi}.
\end{equation}

\begin{figure*}[hbtp]
   \centering
 	\includegraphics[height=1.5in]{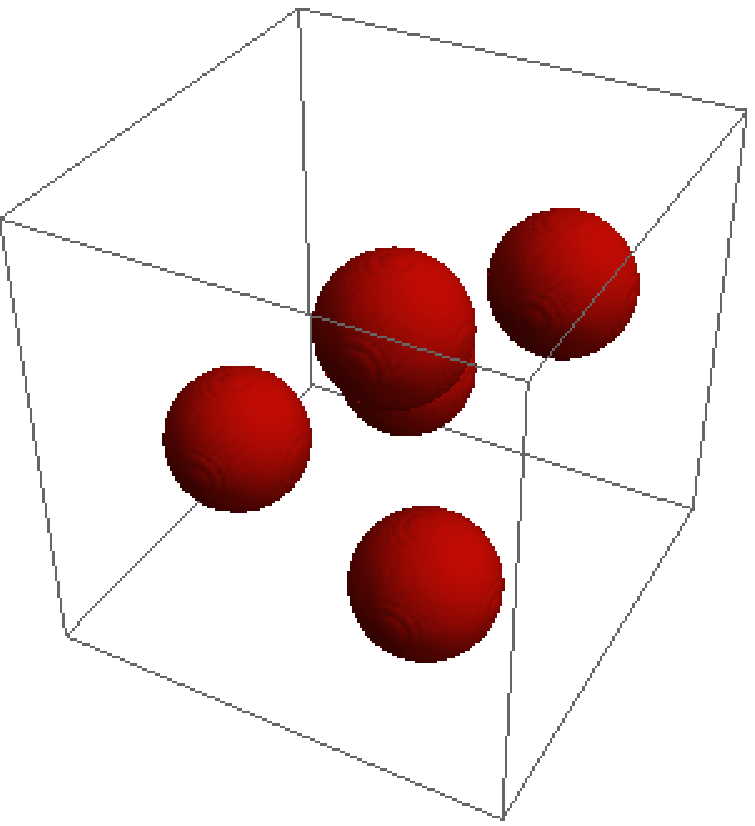}\includegraphics[height=1.5in]{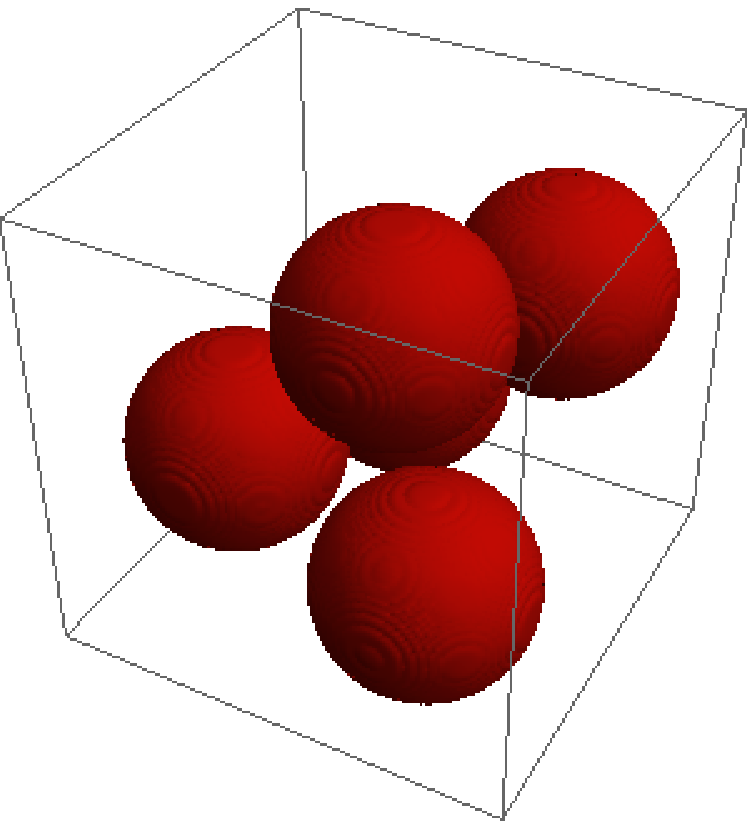}\includegraphics[height=1.5in]{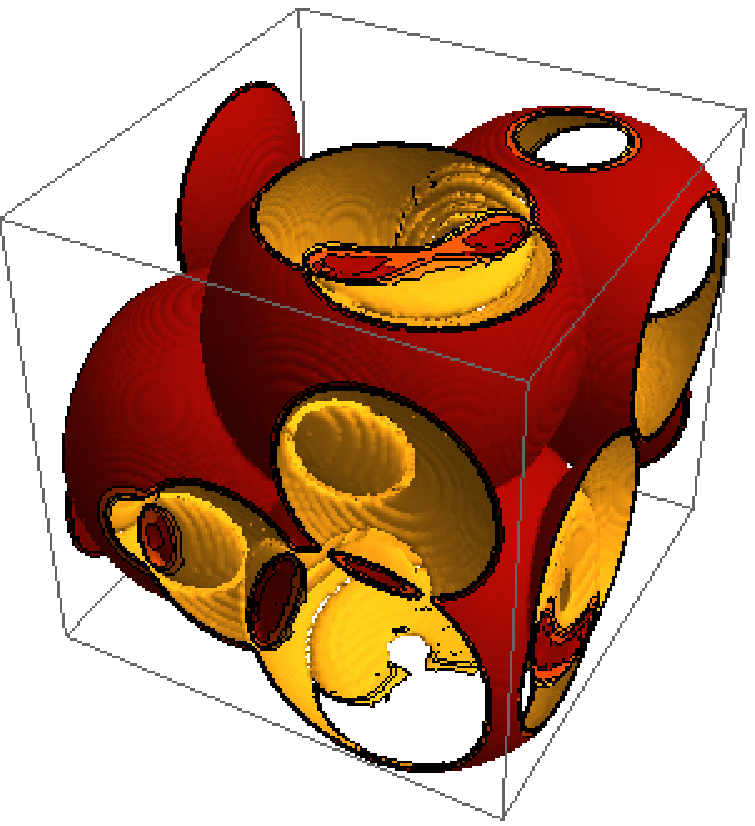}\includegraphics[height=1.5in]{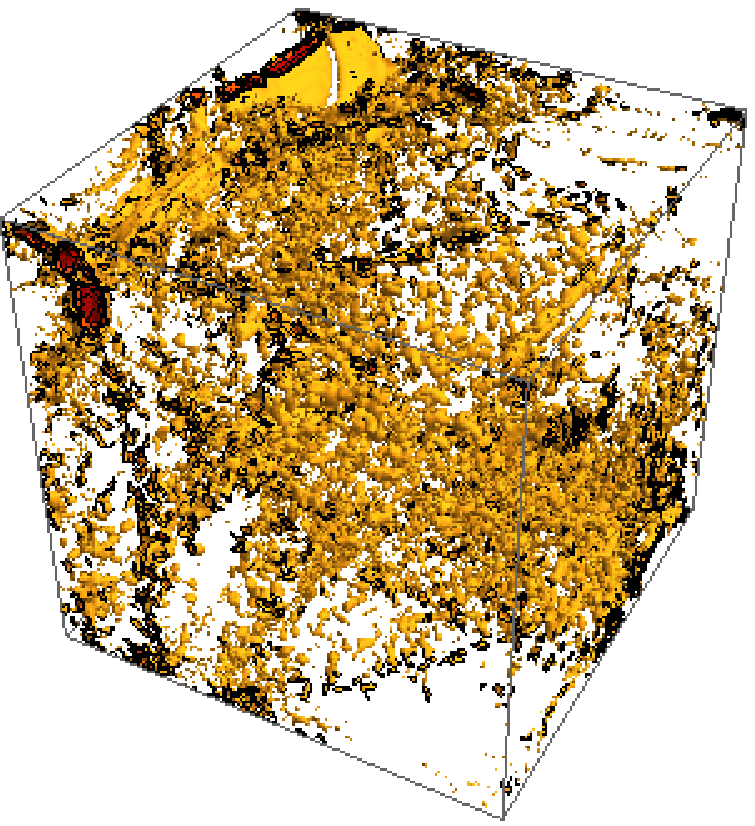}% requires the graphicx package
   \caption{Four time-slices of a first order phase transition when the energy density of the Universe was $\rho\approx (10^{-4}\,m_{\rm pl})^4$.  The first slice shows the nucleation of five bubbles at $\tau=0$, followed by a slice taken as the bubbles initially collide ($\tau \approx 0.5 \beta^{-1}$), a slice at the end of the phase transition ($\tau \approx \beta^{-1}$) and finally a slice when $\tau \approx 2\beta^{-1}$. Contours are drawn at $\phi=-0.83\phi_0$ (gold) and $\phi=0.83\phi_0$ (red) to guide the eye.}
   \label{evolution}
\end{figure*}

The field obeys the usual Klein-Gordon equation in an expanding background,
\begin{equation}
\label{kleingordon}
\ddot{\phi} + 3H \dot{\phi} - \frac{\nabla^2\phi}{a^2} + \frac{\partial V(\phi)}{\partial \phi} = 0,
\end{equation}
which is coupled to Friedmann's equation,
\begin{equation}
H^2 = \left(\frac{\dot{a}}{a}\right)^2 = \frac{8\pi}{3m_{\rm{pl}}^2}\rho = \frac{8\pi}{3m_{\rm{pl}}^2}\frac{\rho_*}{a^4}
\end{equation}
where the final expression is satisfied since we assume the Universe is radiation-dominated throughout the phase transition.  Our choice of $\xi < 1$ allows for this assumption to be valid.

We consider the case where $\xi = 0.25$ and $n$ bubbles nucleate simultaneously per Hubble volume, $H_*^{-3}$; our simulations will involve a smaller volume, $V = H^{-3}_*/8$ with $N=n/8$ bubbles.  At the time of nucleation, each bubble has the field profile of the Coleman bounce \cite{Coleman:1980aw},
\begin{equation}
\phi (x,t_*) =  \phi_0 \tanh\left(\frac{\phi_0\sqrt{\lambda}}{2}(r-r_*)\right)
\end{equation}
where
\begin{equation}
r_*=(\epsilon \phi_0 \sqrt{\lambda})^{-1}
\end{equation}
is the initial bubble radius.  Although the tunneling rate, $\beta$, is set by the parameters of the model ($\lambda$, $\phi_0$ and $\epsilon$) \cite{Coleman:1977py}, we consider a toy model in which the tunneling rate is a free parameter, or rather, that the number of bubbles per Hubble volume, $N$, is a free parameter.

\section {Numerical Simulations}  

The classical production of gravitational radiation from lattice simulations of scalar fields has received a lot of attention over the last decade~\cite{Easther:2006gt,Easther:2006vd,Felder:2006cc,GarciaBellido:2007dg,Easther:2007vj,Dufaux:2007pt,GarciaBellido:2007af,Price:2008hq,Dufaux:2008dn}, particularly in studying the production of gravitational radiation from preheating.  The majority of this analysis has employed {\sc LatticeEasy}~\cite{Felder:2000hq} to simulate scalar fields in an expanding background, although different authors have chosen different methods of calculating the power in gravity waves.  We continue to use a modified version of {\sc LatticeEasy} for our scalar evolution.

In general, we will set the energy scale of the simulations, $\mu$, to $10^{-4}\, m_{\rm pl}$, $10^{-5}\, m_{\rm pl}$, and $10^{-6} \, m_{\rm pl}$. The other parameters in the potential are constrained by $\xi$ and by the ratio of the bubble radius, $r_*$, to the initial Hubble length, $H_*^{-1}$.  We chose $r_*H_*^{-1}=0.07$ so that up to five bubbles can be easily nucleated on the lattice, where the initial length of each side is $H_*^{-1}/2$.  We will always use a $3$-dimensional lattice with finite-time differencing and we define program time $\tau = t-t_*$.

\begin{figure}[htbp]
   \centering
 	\includegraphics[width=3.25in]{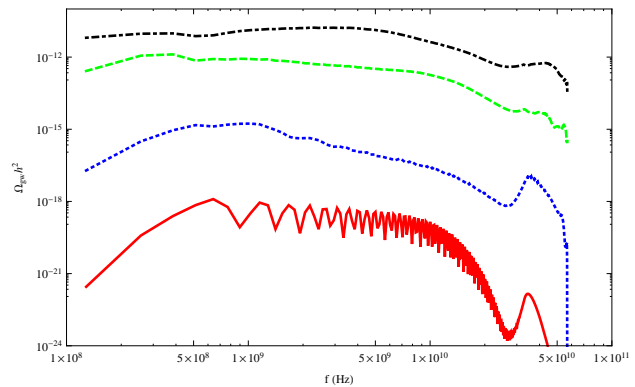}% requires the graphicx package
   \caption{Gravitational wave spectra from a first-order phase transition when the energy density of the Universe was $\rho\approx (10^{-4}\,m_{\rm pl})^4$: the spectrum immediately after the nucleation of five bubbles (red, solid), at $\tau \approx 0.5 \beta^{-1}$ (blue, dotted), when $\tau \approx \beta^{-1}$(green, dashed), and at $\tau \approx 2\beta^{-1}$ (black, dot-dashed).  The bump at high frequencies is a numerical artifact.}
   \label{evolutiongw}
\end{figure}

We begin each simulation by randomly placing $N$ bubbles on the grid.  The average spacing between the bubbles, $L_*$, is related to the nucleation rate $\beta$ and can be calculated by \cite{Chandrasekhar:1943ws}
\begin{equation}
L_* = \beta^{-1} \approx .55396 n^{-1/3} H_*^{-1},
\end{equation}
where $n=8N$ is the number of bubbles per Hubble volume.  After initializing the bubbles of true vacuum, we evolve the field in a radiation-dominated expanding background until a final time $t_f\approx H_*^{-1}/2$, taking $10^4$ time steps on a $512^3$ grid. Over the course of the simulation, the size of the universe increases by a factor of $a=1.5$. Figs.~\ref{evolution} and \ref{evolutiongw} show the time evolution of the field and gravitational wave spectrum for a simulation with $\rho_*=(10^{-4}\,m_{\rm pl})^4$ and $N=5$ bubbles ($n=40$).
The fields then evolve according to Eq.~\ref{kleingordon} until $\tau \approx 2.5 \beta^{-1}$.

We use the algorithm of \cite{Easther:2006vd,Easther:2007vj} to calculate the power spectrum of gravitational radiation produced in this phase transition. The classical metric perturbations, $h_{ij}$, written in synchronous gauge, are
\begin{equation}
\label{sync}
ds^2 = dt^2 - a^2(t)\left[\delta_{ij} + h_{ij}\right]dx^idx^j.
\end{equation}
When we impose the transverse-traceless conditions, ${h^{TT}}_i^i = 0$ and ${h^{TT}}^{ij}, j =0$, the metric perturbations are subject to the equation of motion,
\begin{equation}
\label{hijeom}
\ddot{h}^{TT}_{ij} + 3H \dot{h}^{TT}_{ij} - \frac{\nabla^2h^{TT}_{ij}}{a^2}  = \frac{16 \pi}{m_{pl}^2}S^{TT}_{ij}.
\end{equation}
The source of this wave equation is the transverse-traceless projection of the anisotropic stress tensor, 
\begin{equation}
\label{aniso}
S_{ij} = T_{ij} - \frac{\eta_{ij}}{3}T.
\end{equation}
We evolve the metric perturbations alongside the scalar fields;  this allows us to track the power in gravitational radiation at any point during the simulation and pin-point from where that radiation emerges.  The stress-energy associated with metric perturbations is \cite{Misner:1974qy}, 
\begin{equation} 
T^{\rm
gw}_{\mu\nu} = \frac{m_{\rm pl}^2}{32\pi} \left\langle h_{ij,\mu}h^{ij}_{\,\,\,,\nu}\right\rangle,
\end{equation} 
where the brackets denote a spatial average over several wavelengths.  The $00$ component is the energy density,
\begin{equation} 
\rho_{\rm{gw}} = \frac{m_{\rm pl}^2}{32
\pi} \sum_{i,j} \left\langle\dot{h}^2_{ij}\right\rangle, \label{gwdensity} 
\end{equation}
which can be converted to momentum space by use of Parseval's theorem (see
\cite{Easther:2007vj}) 
\begin{equation} 
\rho_{\rm gw} =\frac{m_{\rm pl}^2}{32\pi}
\frac{1}{V}\sum_{i,j}\int d^3\mathbf{k}\,\,
\Bigl|\dot{h}_{ij}(t,\mathbf{k})\Bigr|^2, \label{omega0} 
\end{equation} 
where $V$ is the comoving volume over which the spatial average is being performed. We then write 
\begin{equation} 
\frac{d\rho_{\rm gw}}{d\ln k} = \frac{m_{\rm pl}^2k^3}{32\pi}
\frac{1}{V} \sum_{i,j} \int d\Omega\, \Bigl| \dot{h}_{ij}^{\rm
TT}(t,\mathbf{k}) \Bigr|^2. \label{omega} 
\end{equation} 
This corresponds to a present-day amplitude and frequency by ~\cite{Easther:2007vj,Price:2008hq},
\begin{equation} 
\Omega_{\rm gw,0}h^2 = \Omega_{\rm rad,0}h^2
\Biggl(\frac{g_0}{g_*}\Biggr)^{1/3} \frac{1}{\rho_{\rm tot, e}}\frac{d\rho_{\rm
gw,e}}{d\ln k}, \label{omega1} 
\end{equation} 
and 
\begin{equation}
f = 6 \times 10^{10} \frac{k}{\sqrt{m_{\rm{pl}}H_e}} \,{\rm Hz}
\label{freqtransfer}
\end{equation}
where the subscript $0$ indicates quantities defined today and $e$ can denote any time during (or at the end of) the simulation.  We also keep the convention that $h$ absorbs the uncertainty in the present value of the
Hubble parameter.  The quantity $\Omega_{\rm rad,0}$ is the current fraction of the energy
density in the form of radiation, and $\rho_{\rm tot, e}$ is the total energy
density at $t_e$. The ratio $g_0/g_*$ compares the number
of degrees of freedom today to the number of degrees of freedom at
matter/radiation equality.  We approximate $g_0/g_*=1/100$.
Fig.~\ref{evolutiongw} shows four time slices (corresponding to the same times as in Fig.~\ref{evolution}) of the gravitational radiation produced in a simulation with $\rho_*=(10^{-4}\,m_{\rm pl})^4$ and $N=5$ bubbles ($n=40$).

\section{Results} 

\begin{figure}[htbp]
   \centering
   \includegraphics[width=3.25in]{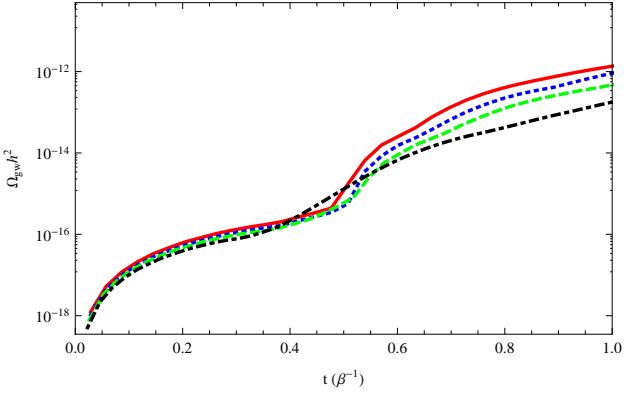} % requires the graphicx package
   \caption{The maximum intensity of the gravitational wave spectrum for a $\mu = 10^{-4}\,m_{\rm pl}$ simulation initialized with 40 (red, solid), 32 (blue, dotted), 24 (green, dashed), or 16 (black, dot-dashed) bubbles per Hubble volume, $\tau<\beta^{-1}$.}
   \label{earlyheightbinv}
\end{figure}

As the bubbles expand and begin to collide, individual collisions are seen to contribute to the power spectrum. These are no longer apparent by $\tau \approx \beta^{-1}/2$, after which time $h^2 \Omega_{gw}$ rises more steadily, as shown in Fig. \ref{earlyheightbinv}.  We begin by considering the time interval $0 < \tau < \beta^{-1}$. 

\subsection{Collisions: $0 < \tau < \beta^{-1}$ }
\begin{figure}[htbp]
   \centering
   \includegraphics[width=3.25in]{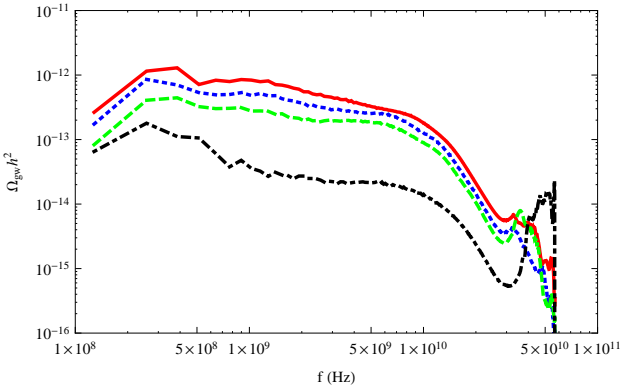} % requires the graphicx package
   \caption{The present-day gravitational wave spectrum produced by time $\tau=\beta^{-1}$.  This is for a simulation with $\mu=10^{-4}\,m_{\rm pl}$ and $n=16$ bubbles per Hubble volume (red, solid), $n=24$ (blue, dotted), $n=32$  (green, dashed), and $n=40$ (black, dot-dashed). The bump at high frequencies is a numerical artifact.}
   \label{separations1}
\end{figure}

During this stage, the peak frequency of the gravitational radiation should correspond to the scale $a_e\beta^{-1}=L_*$ \cite{Kamionkowski:1993fg}, the mean distance between bubbles on the initial slice.  This corresponds to a physical wavenumber $k_{\rm{phys}} = 2 \pi L_*^{-1} \approx 11.34 n^{1/3} H_* $. 
%$L_0 = 0.174 H_0^{-1}$, when $n=4$, so $k_{max}=36 H_0$. 
The associated frequency observed at the present day is given by Eq.~\ref{freqtransfer},
where $H_e$ is the Hubble constant at the time when we calculate the spectrum; $H_e = \frac{H_0}{a^2_e} \sim H_0$, where $a_e$ is the scale factor at the time when we take the spectrum. Putting this together, we expect the peak frequency to occur at
\begin{equation}
f_{\rm peak} \approx 6.8\times 10^{11}n^{1/3}\sqrt{\frac{H_0}{m_{\rm pl}}} \,{\rm Hz} \approx 1.16\times10^{12}\mu\,n^{1/3}\,{\rm Hz}
\end{equation}
This corresponds to a peak around $f_{\rm peak}\sim 10^8\,{\rm Hz}$ when $\mu=10^{-4}\, m_{\rm{pl}}$ and $n\sim10$. These numbers are consistent with the location of a low-frequency peak visible in the spectrum at early times; see Fig.~\ref{separations1}.

The frequency at which the maximum intensity of gravity waves is expected to be found is also given in \cite{Kamionkowski:1993fg} as
\begin{equation}
f_{\rm{peak}} \approx 5.2\times10^{-8}\,{\rm Hz}\left(\frac{\beta}{H_*}\right)\left(\frac{T_*}{1\,{\rm GeV}}\right)\left(\frac{g_*}{100}\right)^{1/6}
\end{equation}
where $g_*$ is the number of ultra-relativistic degrees of freedom at the time of the transition, $\beta$ is, again, the nucleation rate, and $T_*=\mu$ is the energy density when the phase transition occurs. Taking $g_*=400$,
\begin{equation}
f_{\rm{peak}}=1.18 n^{1/3} \mu \times 10^{12},
\end{equation}
which predicts the spectrum to peak at frequencies three or four times $\mu \times 10^{12}$, depending on the number of bubbles. Fig.~\ref{separations1} confirms that for $\mu=10^{-4}$, peak frequencies occur around this value.

A more recent calculation of the peak frequency \cite{Huber:2008hg} predicts
\begin{equation}
f=16.5 \times 10^{-6} \, \rm{Hz} \left( \frac{f_*}{\beta} \right)\left(\frac{\beta}{H_*}\right)\left(\frac{T_*}{100\,{\rm GeV}}\right)\left(\frac{g_*}{100}\right)^{1/6}
\end{equation}
where we use the same values as before and approximate $f_*/\beta$ by
\begin{equation}
\frac{f_*}{\beta}=\frac{0.62}{1.8-0.1v+v^2}=0.22963
\end{equation}
when the bubble wall velocity $v \approx 1$ (the scalar bubbles in our simulation accelerate quickly to $v\approx 1$). This prediction of the frequency simplifies to
\begin{equation}
f_{\rm{peak}}=8.60 n^{1/3} \mu \times 10^{11},
\end{equation}
which also predicts a few times $10^{8}$, also consistent with Fig.~\ref{separations1}.

In addition to calculating the peak frequency, \cite{Kamionkowski:1993fg} also gives the fraction of critical density found in gravity waves as
\begin{equation}
\begin{split}
\Omega_{gw}h^2 =& 1.1 \times 10^{-6} \kappa^2 \left(\frac{H_*}{\beta}\right)^2\left(\frac{\alpha}{1+\alpha}\right)^2\\
&\left(\frac{v^3}{0.24+v^3}\right)\left(\frac{100}{g_*}\right)^{1/3}
\end{split}
\end{equation}
where the efficiency factor $\kappa$ measures energy lost to the motion of a coupled fluid. We set $\kappa = 1$, as our simulations do not include a fluid. The fraction $H_*/\beta$ is the mean initial bubble separation as a fraction of the Hubble distance,  $v$ is the velocity of the bubble walls, and $g_*$ is, again, the number of ultra-relativistic degrees of freedom.  In our simulations we set $\alpha = 1/3$.  Assuming $v=1$ and $g_* = 400$, the expected intensity of the spectrum reduces to 
\begin{equation}
\Omega_{gw}h^2=1.42n^{-2/3} \times 10^{-9}.
\end{equation}
This estimate varies between 90 and 1250 times greater than simulation results at $\tau = \beta^{-1}$, becoming more consistent in the large $n$ limit.  
The more recent prediction \cite{Huber:2008hg} of intensity as
\begin{equation}
\begin{split}
\Omega_{gw}h^2 = &1.67 \times 10^{-5} \left(\frac{0.11v^3}{0.42+v^2}\right) \kappa^2 \left(\frac{H_*}{\beta}\right)^2 \\
 &\left(\frac{\alpha}{1+\alpha}\right)^2\left(\frac{100}{g_*}\right)^{1/3} \\
= & 2.07n^{-2/3} \times10^{-9}
\end{split}
\end{equation} 
falls slightly farther from our results. There are two main factors that describe the discrepancies:  (1) we have implemented an expanding background in which Hubble friction depletes the energy of the source by a small fraction and (2) our models realistically thicken the bubble walls. ``Thick" walls prolong collisions and dilute the gradient terms that source strong gravitational waves.  At the same time the inclusion of these effects strengthens the validity of the current model.

\subsection{Coalescence: $\beta^{-1} < \tau < 2 \beta^{-1}$:}

\begin{figure}[htbp]
   \centering
   \includegraphics[width=3.25in]{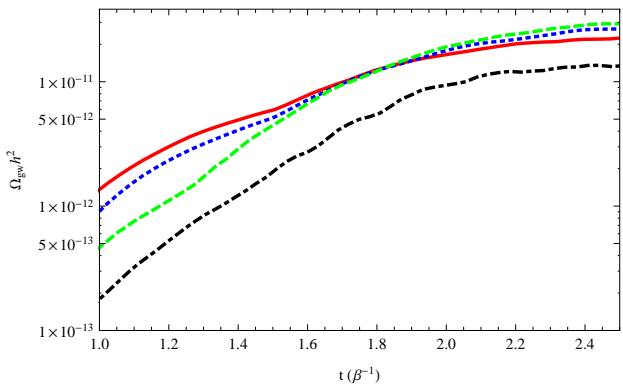} % requires the graphicx package
   \caption{The maximum intensity of the gravitational wave spectrum for a $\mu = 10^{-4}\,m_{\rm pl}$ simulation initialized with 40 (red, solid), 32 (blue, dotted), 24 (green, dashed), or 16 (black, dot-dashed) bubbles per Hubble volume, $\beta^{-1}<\tau<2.5\beta^{-1}$.}
   \label{lateheightbinv}
\end{figure}

Although most of the volume of the simulations is in the true minimum by $\tau=\beta^{-1}$, there is still a lot of kinetic, gradient and even potential energy in the fields.  The phase transition is, more or less, complete, but the production of gravitational radiation has not ceased.  Fig.~\ref{lateheightbinv} shows how the peak amplitude rises from $\tau=\beta^{-1}$ until $\tau= 2.5\beta^{-1}$.

We expect that a period of turbulence can increase the magnitude of the spectrum by several orders of magnitude, e.g.  \cite{Kamionkowski:1993fg,Caprini:2006jb,Gogoberidze:2007an,Caprini:2007xq,Kahniashvili:2008pf,Caprini:2009fx}. One of these estimates  \cite{Kamionkowski:1993fg} say that the intensity of gravitational radiation after turbulence is predicted to be
\begin{equation}
\Omega_{gw}h^2 = 10^{-5}\left(\frac{H_0}{\beta}\right)^2v v_0^6\left(\frac{100}{g_*}\right)^{1/3},
\end{equation}
which reduces to
\begin{equation}
\Omega_{gw}h^2=2.55n^{-2/3} \times 10^{-7}.
\end{equation}
Thus, the intensity of the gravitational wave spectrum is expected to be around order $10^{-8}$ if we take $v_0 \approx 1$.
We can do slightly better if we try to assign a sound speed, $v_0$, by estimating the speed of perturbations in the true vacuum.  Near $\phi = -\phi_0$, the effective mass of the field, $m^2_{\rm eff} = \lambda \phi_0^2$, and
\begin{equation}
v_0 = \frac{\partial \omega}{\partial k} = \sqrt{\frac{1}{1+\lambda \phi_0^2/k^2}}.
\end{equation}
This ranges between $10^{-2}$ for low frequency modes, $k_{\rm low} \sim H_* \approx \sqrt{\lambda}\phi_0/500$, and almost 1 for higher frequency modes, $k\sim\sqrt{\lambda}\phi_0$.   The final amplitudes that we present here, see Fig.~\ref{lateheightbinv}, are some three orders of magnitude lower, but this is understandable as there is no turbulence {\sl per se} in our simulation.  

There is, however, significant post-collisionary amplification of the gravitational wave spectrum.  This period of {\sl coalescence} after $\tau = \beta^{-1}$ of the simulation, {\sl amplifies the spectrum by more than an order of magnitude}, depositing energy in higher frequency modes.

\begin{figure}[htbp]
   \centering
   \includegraphics[width=3.25in]{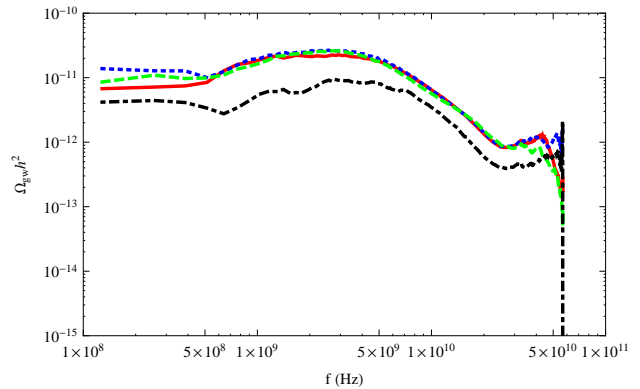} % requires the graphicx package
   \caption{The present-day gravitational wave spectrum at $t=2\beta^{-1}$, nearing the end of the coalescence phase.  This is for a simulation with $\rho_*=(10^{-4}\,m_{\rm pl})^4$ and 16 bubbles per Hubble volume $H_*^{-3}$ (red, solid), 24 bubbles per Hubble volume (blue, dotted), 32 bubbles per Hubble volume (green, dashed), 40 bubbles per Hubble volume (black, dot-dashed).  The bump at high frequencies is a numerical artifact.}
   \label{separations2}
\end{figure}

The frequency of the turbulence peak is \cite{Kamionkowski:1993fg}
\begin{equation}
\begin{split}
f_{\rm{peak}} &\simeq 2.6\times10^{-8}\,{\rm Hz}\,v_0v^{-1}\left(\frac{\beta}{H_*}\right)\left(\frac{T_*}{1\,{\rm GeV}}\right)\left(\frac{g_*}{100}\right)^{1/6}\\
&=5.91 \times 10^{11}n^{1/3} \mu \times 10^{11}
\end{split}
\end{equation}
This predicts the peak to shift downward during turbulence, while we see a higher-frequency peak at the end of the simulation, with frequency $f \sim \mathcal{O}(\mu) \times 10^{13}$.  This peak comes from the amplification of higher frequency modes during the coalescence period.  Fig.~\ref{separations2} shows the integrated gravitational wave spectrum for $\mu = 10^{-4}$ after this period.  In this plot the peak at higher frequencies is quite apparent at $\tau=2.5\beta^{-1}$.

It should be noted that Fig.~\ref{separations2} does not explicitly show the $k^3$ low-frequency tail of the gravitational wave spectrum.  This is due to the lack of resolution at the relevant scales;  the longest wavelength that we can resolve is well within the horizon at the time of the phase transition.  The leftmost few points in all of our spectra are averaged only over a small number of modes and should not be used to extrapolate to very low frequencies.

\begin{figure}[htbp]
   \centering
   \includegraphics[width=3.25in]{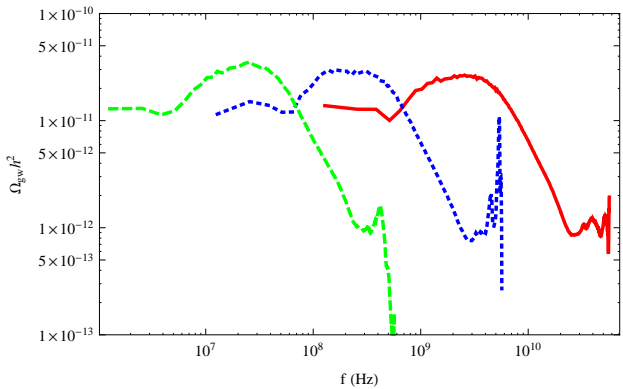} % requires the graphicx package
   \caption{The present-day gravitational wave spectra at $\tau=2.5\beta^{-1}$ for cases with 32 bubbles per Hubble volume at three energy scales: $\mu = 10^{-6}\,m_{\rm pl}$ (green, dashed), $10^{-5}\,m_{\rm pl}$ (blue, dashed) and $10^{-4}\,m_{\rm pl}$ (red, solid). The bump at high frequencies is a numerical artifact. }
   \label{scales}
\end{figure}

Lastly, it's important to check how the spectrum varies with energy scale.  In Fig.~\ref{scales} we vary the energy scale, $\mu$, between $10^{-6}\,m_{\rm pl}$ and $10^{-4}\,m_{\rm pl}$.  The amplitude of the gravity wave spectrum should be independent of scale of the simulation, $\mu$ (also $T_*$ in \cite{Kamionkowski:1993fg,Huber:2008hg} among others), and should only depend on $\beta^{-1}$ and dynamical factors.  Indeed, we recover the scale-independent behavior for the three orders of magnitude that we test.
We anticipate that our simulations would continue to produce similar spectra even at much lower energy scales (modified only slightly when the number of ultra-relativistic degrees of freedom, $g_e$, decreases).

\section{Discussion}

Gravitational radiation should be the most obvious relics of first-order phase transitions that may have existed in the history of the Universe.  To the best of our knowledge, these results represent the first 3-dimensional simulations of first-order cosmological phase transitions and the highest-resolution lattice gravitational wave predictions to date.   

Using these simulations, without a coupled fluid and with only scalar degrees of freedom, we have confirmed previous analytic and numerical simulations of gravity waves from first-order processes, and have reproduced the predicted scalings of the results both in frequency and amplitude.  We had identified the two relevant stages of the process: (1) the stage during which the bubbles collide and (2) a coalescence phase during with the field settles into the true vacuum.  During the first stage, we precisely reproduce the location of the peak of gravitational radiation from previous estimates.  The amplitude at this time is lower than expected from these estimates, due to the inclusion of friction and by realistically ``thickening" the walls.  

Primarily, though, we have discovered that a coalescence phase following the phase transition amplifies the gravitational wave signal for a first-order phase transition by about an order of magnitude and increases the peak frequency by about a decade.  This is most likely a consequence of the persistence of energy in domain walls, even after regions have collided, e.g. \cite{Hawking:1982ga}, residual anisotropic stress-energy produced as the Universe relaxes to a thermal state.  This compensates for the lack of power in gravity waves at $t\approx \beta^{-1}$.

The electroweak phase transition will occur at a much lower frequency than those presented here.  If we estimate this energy scale as $\sim 200\,{\rm GeV}$, then we expect the peak frequency of gravitational radiation from the phase transition to occur at a few times $10^{-5}\,{\rm Hz}$, assuming $\beta/H_*\approx 5$ as we have here.  Coalescence will both amplify the signal and raise the peak frequency about an order of magnitude.  We believe that this is an important effect to be considered in the next generation of detector experiments.

We intend to follow up on this model by adding dynamical fluids to our 3-dimensional simulations.  The evolution of fluids alongside our scalar fields will allow us to confirm the parametric dependence of the gravitational wave signal on the terminal velocity of the bubble walls, check for the amplification of the signal due to the existence of turbulence and confirm that a coalescence phase exists in  the presence of a viscous fluid.

\section{Acknowledgments}  We thank Andrew Tolley  and Eric Greenwood for useful discussions, and Gary Felder for the use of {\sc LatticeEasy}.  We thank Eugene Lim for extremely useful insights on the numerical analysis conducted here.  HLC and JTG are supported by the National Science Foundation, PHY-1068080, and a Cottrell College Science Award from the Research Corporation for Science Advancement.

\end{document}